\documentclass[]{ppmo}                  
\input{epsf.sty}                        
\input{psfig.sty}                       

\newcommand{\micron}{$\mu$m\ }
\newcommand{\micronns}{$\mu$m}

\newcommand{\lsunns}{\hbox{{\it L}$_\odot$}}

\newcommand{\spitzer}{{\it Spitzer}\ }

\begin{document}

\title{{\it Spitzer} Imaging of nearby ULIRGs and their Progeny:
Fine-Structure Elliptical Galaxies}

   \volnopage{Vol.0 (2005) No.0, 000--000}      
   \setcounter{page}{1}          
   \baselineskip=5mm             

\author{Jason~A.~Surace
\inst{1}, in collaboration with\\
(ULIRGs): Z.~Wang\inst{2}, S.~Willner\inst{2}, H.~Smith\inst{2},
J.~Pipher\inst{3}, W.~Forrest\inst{3}, G.~Fazio\inst{2} \\
(Ellipticals): J.~Howell\inst{4}, A.~Evans\inst{5}, J.~Hibbard\inst{6}, L.~Yan\inst{1}, F.~Marleau\inst{1}}

\institute{Spitzer Science Center, MS 220-6, Caltech, Pasadena, CA,
91125\\
\and Harvard Smithsonian Center for Astrophysics, 60 Garden Street, Cambridge, MA
02138
\\
\and Department of Physics and Astronomy, University of Rochester, Rochester, NY
14627\\
\and Infrared Processing and Analysis Center, MS 100-22, Caltech, Pasadena, CA,
91125\\
\and Department of Physics and Astronomy, Stony Brook University, Stony Brook, NY,
11794-3800\\
\and National Radio Astronomy Observatory, 520 Edgemont Road, Charlottesville, VA 22903
}

\date{Received~~2001 month day; accepted~~2001~~month day}

\abstract{ We present results from two mid-infrared imaging programs
of ultraluminous infrared galaxies (ULIRGs) and fine-structure
elliptical galaxies.  The former are known to nearly all be recent
mergers between two gas-rich spiral galaxies, while the latter are
also believed to be even more aged merger remnants.  An examination of
these two classes of objects, which may represent different stages of
the same putative merger sequence, should reveal similarities in the
distribution of their stellar populations and dust content consistent
with that expected for time-evolution of the merger.  The data reveal
for the first time extended dust emission in the ULIRG galaxy bodies
and along their tidal features.  However, contrary to expectation, we
find that the vast majority of the fine-structure elliptical galaxies
lack such structured emission.  This likely results from the optical
selection of the elliptical sample, resulting in a population that
reflects the IR-activity (or lack thereof) of optically selected
interacting galaxies.  Alternately, this may reflect an evolutionary
process in the distribution of the dust content of the galaxy bodies.
}

   \authorrunning{Surace et al.}            
   \titlerunning{Spitzer ULIRGs and Fine-Structure Ellipticals}  
   
   \setlength\baselineskip {5mm}             

\maketitle

\section{Introduction}           
\label{sect:intro}

Ultraluminous Infrared Galaxies (ULIRGs) as a class are characterized
by infrared luminosities greater than 10$^{12}$ \lsunns, equivalent to
the luminosity of classical optically selected quasars.  Based on
extensive ground and space-based imaging, nearly all are known to be
advanced merger remnants resulting from collisions of gas-rich
galaxies.  Dynamics of the collision result in a rearrangement of the
gas, dust, and stellar content of the galaxies, a process well-modeled
by n-body simulations.  The rapid inflow of gas deep into the merger
core results in a powerful burst of star formation, and/or the fueling
of an active nucleus, the latter being known to exist in roughly
one-quarter of ULIRGs.  Arguments based on their space densities,
which are similar to QSOs, led to the idea that they were related
to the QSO population.  It has thus been postulated that the ULIRGs
are the progenitors of both QSOs and elliptical galaxies.  Although
their local space density is relatively low, it undergoes a rapid
increase with (1+$z$).  Because ULIRGs were much more common in the
past, understanding the characteristics of local merger-driven extreme
starbursts and their evolution with time is critical to understanding
their counterparts in the early universe during the time period when
most galaxy formation occurred and later evolved into the current
galaxy population.

Local samples of about 20 ULIRG systems (at $z<0.15$, the closest and
most amenable to study) have been subjected to a great barrage of
observations with a wide array of telescopes and at nearly all
wavelengths.  However, results from these observations have been
highly conflicted.  Extensive imaging campaigns in the UV, optical,
near and mid-infrared during the '90s revealed the composite nature of
the ULIRGs (Surace et al.  1998, 2000ab, Scoville et al.  2000,
Soifer et al.  2000, Goldader et al.  2002).  The ULIRGs have
extremely complex dust absorption, emission, and scattering geometries
resulting from the merger process.  While the outer parts of the
galaxies are relatively optically thin, the inner few kiloparsecs are
generally optically thick, with the cores reaching thousands of
A$_{v}$ and are thus opaque even in the mid-infrared.  At optical and
UV wavelengths the emission is dominated by the old stellar
population, as well as actively star-forming ``super star clusters''
located both in the nuclear regions and along the tidal features.  At
infrared wavelengths the emission is increasingly due to thermal dust
emission arising in a compact core typically less than 100pc in
diameter.  Ground-based mid-infrared observations were able to
constrain a large fraction of the IRAS 12 and 25\micron emission to
this unresolved core.  This explains their confusing properties, which
result from compositing many different, unresolved emission sources.

The Spitzer Space Telescope is especially well-suited for examining
these dust-enshrouded systems.  Infrared observations penetrate the
dust to greater depth.  The IRAC camera has filters that specifically
sample the old stellar photospheric emission, useful for
tracing the underlying stellar distribution, and also the line
emission from polycyclic aromatic hydrocarbons (PAHs), which traces
dust and star formation and which is unusually strong at 8\micron in
ULIRGs.

\section{ULIRGs Observed by Spitzer}
\label{sect:ULIRGs}

As part of the IRAC GTO program, we observed 14 nearby ULIRGs selected
from the well-studied Bright Galaxy and Warm Galaxy Samples of Sanders
et al.  (1988).  The spatial resolution of \spitzer is relatively poor
($\approx$2\arcsec) at IRAC wavelengths compared to the known optical
compact structure in these galaxies.  However, it is sufficient to
resolve the galaxy bodies and extended tidal features, and to separate
the double nuclei known to exist in some systems at optical
wavelengths.  While in principle ground-based observations (e.g. from
Keck) offer several times higher spatial resolution, they lack the
sensitivity to effectively take advantage of it.  The extended, and
hence potentially resolvable, emission is too faint to be readily
detected from the ground. {\it Spitzer}, however, has immensely greater
sensitivity to low surface brightness emission, and is the only
current means of detecting the extended galaxy bodies.

\begin{figure}
    \plotone{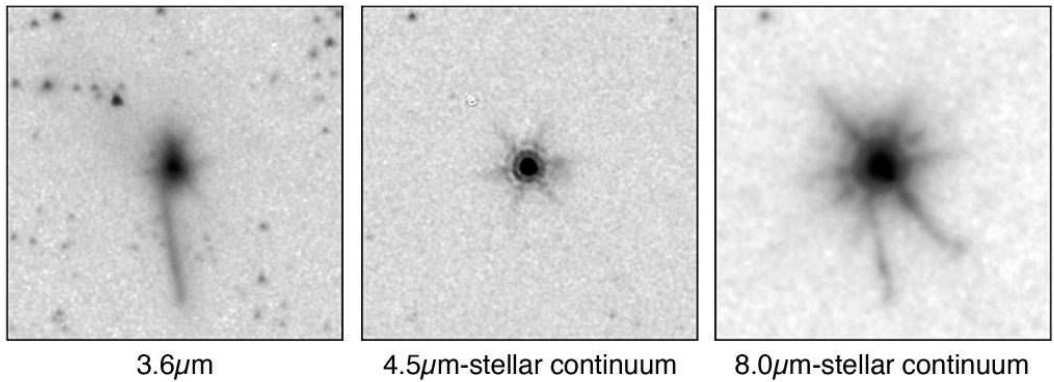}
    \caption{{\it Spitzer} images of the ULIRG Mrk~273. At left is
    the 3.6\micron data, clearly showing the extended galaxy body and 
    long tidal tail extending to the south. In the middle is the
    4.5\micron image, with a model of the stellar continuum
    subtracted. This very effectively subtracts away the galaxy body, 
    showing that the only significant source of IR-excess at this
    wavelength is the compact nuclear core. However, when the same
    operation is performed on the 8\micron data, knotty extended
    excess emission, presumably from PAHs, is shown along the tidal
    tail.
    Note that the linear structure extending NE to SW is an artifact
    of the detector.}
    \end{figure}

We therefore specifically selected systems that had potentially
resolvable optical emission on spatial scales of several arcseconds or
more, and thus excluded some nearby yet compact ULIRGs like
IRAS~01003-2238.  The selected galaxies include all of the canonical
favorites such as Arp~220, Mrk~231, and UGC~5101.  The systems were
mapped over areas several times larger than the extent of the known
optical emission.  Imaging was carried out with IRAC at 3.6, 4.5, 5.8,
and 8\micronns, and with MIPS at 24, 70, and 160\micron in a few
cases.  These \spitzer observations were surprisingly challenging for
several reasons, although primarily because the central cores were so
bright.  The MIPS mid and far-infrared data have not been particularly
enlightening beyond adding an additional photometric data point to the
observed spectral energy distributions (SED), as the complex point
spread function of \spitzer and the extreme surface brightness ratio
between the nuclear core emission and the extended stellar
distribution prevents the detection of the latter.  Similarly, the
IRAC data suffered from significant artifacts associated
with bright point sources.  Additionally, there are significant
photometric uncertainties for small-scale (arcminute) extended
emission.

All of the ULIRGs were well-detected.  Their spatial structure at 3.6
and 4.5\micron is very similar to the near-infrared, unsurprising
since at these wavelengths emission from old stellar photospheres
clearly dominates the galaxy bodies.  In almost all cases an extended,
low surface brightness galaxy body is seen with accompanying tidal
structure.  An unresolved (by {\it Spitzer}) compact core lies at the
galaxy center.  This core is increasingly dominant at longer
wavelengths, reflecting the falling SED of the stars vs.  the rising
SED of the thermal dust emission associated with the core.  Double
nucleus systems appear somewhat more extended than single nucleus
systems and no ``new'' emission beyond the known optical and infrared
extent of the galaxies has been seen.

The ULIRG progenitors are widely believed to be spiral galaxies, which
characteristically have prodigious amounts of dust distributed in
flocculent structures throughout their disks.  During the merger
process it is expected that this dust flows inward toward the galaxy
center.  The extended structure in ULIRGs is thus particularly
interesting as it will reveal whether or not the dust has been
stripped from the galaxy bodies, either due to gravitational inflow or
wind-borne outflow.  In a manner similar to that of Pahre et al.
(2004), we have modeled the underlying stellar distribution based on
the continuum-dominated emission at 3.6\micronns, including the effect
of cross-convolution of the relative beam shape between wavebands.
This is illustrated in Figure 1.  Many of the ULIRGs show clearly
extended emission at 8\micronns.  The most commonly cited ULIRG,
Arp~220, is found to have nearly half of it's total 8\micron emission
arising in an extended structure outside the unresolved nucleus.  This
structure corresponds closely to the known extended CO emission, but
not with the axes of the known gaseous emission line nebula
(superwind).  Little dust is found in the extended tidal tails.
However, many other ULIRGs do show dust in their extended tidal
structures.  The figure of Mrk~273 clearly shows dust at 8\micron
distributed along it's prominent tidal tail.  About 20\% of the total
8\micron flux originates in extended emission outside the nuclear
core.  This correspondence with stellar and tidal structures in
the galaxy bodies strongly suggests that the dust is not entrained in
winds, but is rather either left over from the progenitor galaxies or
was formed in situ by recent star formation.

\section{Fine-Structure Elliptical Galaxies Observed by Spitzer}
\label{sect:Ells}

It is widely believed that galaxy-galaxy major mergers like those that
form the ULIRGs evolve over time into systems resembling elliptical
galaxies.  This argument is grounded theoretically in n-body computer
simulations, and is supported observationally on several grounds
including similarity of their isophotes and stellar velocity fields to
elliptical galaxies (\cite{1990Natur.344..417W}, \cite{2002ApJ...580...73T}).  Most
studies of mergers have focused on relatively early stage mergers
(e.g. the Toomre sequence) or infrared-selected late-stage mergers
(e.g. ULIRGs).  However, there has been very little investigation of
what these merger systems eventually evolve into.  The \spitzer data
has revealed the distribution of stars and dust within the ULIRGs.
Any putative descendants of the ULIRGs should show relics of these
features, as evidenced either by resolved structures in the
observed stellar distribution or in the form of extended dust.

\begin{figure}
    \plotone{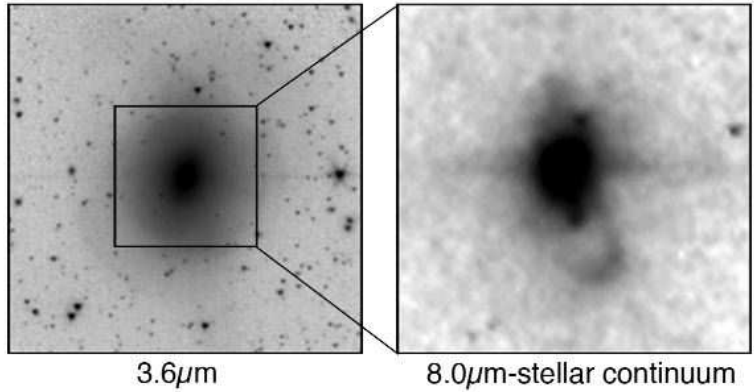} 
    \caption{NGC~5018, an elliptical galaxy
    with complex dust features.  At left is the 3.6\micron IRAC image,
    and at right is the 8\micron image minus a stellar continuum
    model.  The resulting knots and loops are similar to those found
    in ULIRGs.  However, this object is a clear minority in our
    sample.}
    \end{figure}

It is well-known that certain elliptical galaxies exhibit
morphological disturbances, usually referred to as ``fine-structure'',
and which are believed to be evidence of merger activity in the past
few Gyrs.  Examination of such objects can bridge the gap between
late-type mergers such as ULIRGs and the general elliptical galaxy
population.  The new observations can help determine the relation and
timescales between the two classes of objects, as well as reveal
details of the evolution of the dust content in advanced mergers.
As part of a {\it Spitzer} GO-1 program we have investigated a sample
of these fine-structure ellipticals.  The sample is drawn from
\cite{1992AJ....104.1039S}.  The data were processed and analyzed in a
manner similar to that described above for the ULIRGs.

All of the fine-structure ellipticals have clear distortions visible
in the mid-infrared in the distribution of the old stellar population,
primarily consistent with shells and other merger-induced features in
the isophotal profiles.  However, except in a small fraction of
systems (see Figure 2) there is {\it little or no} evidence for widespread systemic
dust emission or high surface brightness features from PAHs at
8\micronns, as are characteristically found in the ULIRGs.  There is
some evidence for smooth color gradients and excess emission in the
galaxy cores.  Unfortunately, interpretation of these gradients is
complicated by photometric uncertainties in the extended emission
resulting from scattering in the IRAC camera.

It is likely that this lack of dust emission results from the sample
selection.  The elliptical galaxies were selected based on optical
features such as shells or dust lanes.  However it is known that while
nearly all IR-active galaxies are interacting, the converse is not
true: most mergers are not particularly IR-active, which probably
reflects the pre-existing richness of the ISM in the merger
progenitors or the interaction geometry.  Most of the ellipticals
selected probably resulted from mergers that were never particularly
IR-active.  The dust features that do exist often appear to arise from
cold, thin dust given that in some cases they are seen in optical
absorption but not in mid-infrared thermal emission.

\section{Future Work}
\label{sect:future}

Studying the dynamics of the dust in merger systems will be fruitful
for understanding the evolution of the SED in ULIRGs and their
possible evolution into QSOs or elliptical galaxies.  The \spitzer
observations are the most sensitive that can be achieved near-term.
Future imaging by JWST will provide a much clearer view of the ULIRG
dust morphology.  Resolved thermal imaging of dust in ellipticals is still in
its infancy - we lack an overall understanding of dust in ellipticals
due to a lack of large samples.  Hopefully this will change over {\it
Spitzer}'s lifetime.  Different approaches to elliptical sample
selection, coupled with ground-based imaging and SEDs derived from
previous infrared space missions, could also advance this
study considerably.

\begin{acknowledgements}
The author was supported by the Jet Propulsion Laboratory, California
Institute of Technology, under contract with NASA. The author also
thanks the National Science Foundation
of China for support to attend the conference ``Extreme Starbursts:
Near and Far''.

\end{acknowledgements}

\label{lastpage}    
    
\end{document}